
\magnification=1200  
\def\newline{\hfill\penalty -10000}  
\def\title #1{\centerline{\rm #1}}
\def\author #1; #2;{\line{} \centerline{#1}\smallskip\centerline{#2}}
\def\abstract #1{\line{} \centerline{ABSTRACT} \line{} #1}
\def\heading #1{\line{}\smallskip \goodbreak \centerline{#1} \line{}}
\newcount\refno \refno=1
\def\refjour #1#2#3#4#5{\noindent    \hangindent=1pc \hangafter=1
 \the\refno.~#1, #2 ${\bf #3}$, #4 (#5). \global\advance\refno by 1\par}
\def\refbookp #1#2#3#4#5{\noindent \hangindent=1pc \hangafter=1
 \the\refno.~#1, #2 (#3, #4), p.~#5.    \global\advance\refno by 1\par}
\def\refbook #1#2#3#4{\noindent      \hangindent=1pc \hangafter=1
 \the\refno.~#1, #2 (#3, #4).           \global\advance\refno by 1\par}
\newcount\equatno \equatno=1
\def\adveqn{(\the\equatno) \global\advance\equatno by 1}

\def\up#1{\leavevmode \raise 0.2ex\hbox{#1}}

\newcount\figno
\figno=0
\def\figure{\global\advance\figno by 1 Figure~\the\figno.~}
%

%
\vsize=8.75truein
\hsize=5.75truein
\hoffset=0.5truein
%
\baselineskip=0.166666truein
%
\parindent=25pt
%
\parskip=0pt
%
\nopagenumbers

%
%
\line{}

\title{
RADIATION FROM VELA-LIKE PULSARS NEAR THE DEATH LINE
}

\author
Vladimir V. Usov \parindent=0pt ;
Dept.~of Physics, Weizmann Institute, Rehovot 76100, Israel ;

\abstract{
Radiation of both the outer gaps and the neutron star surface
is considered for a Vela-like pulsar near the death line.
It is shown that if such a pulsar is close enough to the
death line, its optical, UV and X-ray emission has to increase.
Using results of this consideration, it is argued that Geminga is not
a close relative of Vela-like pulsars. The outer gap model of
Geminga in which the main part of the outer gap volume operates as a
Vela-like generator of $\gamma$-rays is ruled out. A Vela-like
mechanism of $\gamma$-ray generation can operate only in a
small region of the outer gap of Geminga.
The length of this region along
the magnetic field is an order of magnitude smaller than the outer gap
dimensions. In the magnetosphere of Geminga
the main mechanism of $\gamma$-ray generation at $\sim 10 - 10^4$
MeV is curvature radiation and not the
synchrotron radiation as it was assumed for Vela-like pulsars.}

\heading{1. INTRODUCTION}
It has been argued by Holloway (1973), Cheng, Ruderman and
Sutherland (1976) and Krause-Polstorff and
Michel (1985 a,b) that charge deficient
regions ("outer gaps") with a strong electric field $E_\parallel$
along the
magnetic field may exist near the pulsar light cylinder. The
structure of outer gaps and their radiation were considered by
Cheng, Ho and Ruderman (1986 a,b, hereafter CHRa,b).
The CHR "outer gap" model describes the
high frequency (optical, X-ray and $\gamma$-ray)
radiation of the Crab and Vela pulsars fairly well.
\par Outer gaps may act as a particle accelerator in the
pulsar magnetosphere only if the period of the pulsar rotation $P$ is
small enough (CHRb, Chen and Ruderman 1993, hereafter CR).
Ruderman and Cheng (1988, hereafter RC) have predicted
that when a Vela-like pulsar in the process of deceleration of its
rotation approaches to the death line, the pulsar luminosity in
$\gamma$-rays increases and may be as high as the total
spin-down power of the neutron star. Below it is shown that
not only the $\gamma$-ray emission of a Vela-like pulsar has to
increase near the death line but its optical, UV and X-ray
emission has to increase as well.
\par Nonthermal radiation from Vela-like pulsars near the
death line is considered in Section 2. For such a pulsar
the lower limit on the thermal X-ray emission of the
neutron star surface is given in Section 3. Geminga is considered
as a candidate for a Vela-like pulsar in Section 4. The
discussion and summary of conclusions are presented in
Section 5.

\heading{2. OPTICAL AND UV EMISSION FROM OUTER GAPS NEAR THE DEATH LINE}
In the outer gap model of CHRb the radiation of a Vela-like pulsar
is generated by means of the following long sequence of processes:
1. Acceleration of primary particles by strong electric
field $E_\parallel$; 2. Generation of primary $\gamma$-rays through
inverse Compton scattering on the tertiary photons; 3. Creation of
secondary electron-positron pairs in collisions of primary
$\gamma$-rays with the same tertiary photons; 4. Generation of
secondary $\gamma$-ray and X-ray emission by secondary pairs
through synchrotron  radiation; 5. Creation of low energy (tertiary)
pairs in collisions of the secondary hard $\gamma$-rays with the
secondary soft X-rays; 6. Generation of the tertiary synchrotron
radiation by the tertiary pairs which initiates the entire series
of pair production processes in the outer gap.
\bigskip
2.1 {\it Tertiary photons: typical frequency and luminosity}
\medskip
The typical frequency of the tertiary photons is (CHRb; CR)
$$\omega_s \simeq {1\over \omega _{_B}^3}\left(
{mc^3\Omega\over e^2}\right)^2 \simeq 10^{16}
\left({\bar B\over 10^4\, {\rm Gauss}}\right)^{-3}\left({\Omega
\over 70\,\,{\rm s}^{-1}}\right)^2\,\,
{\rm s}^{-1}\,, \eqno (1)$$
where $\Omega = 2\pi/P$ is an angular velocity of the neutron
star rotation, $\omega_{_B} =e\bar B/mc$ and $\bar B$ is
a gap-averaged magnetic field. This equation is valid if $\omega
_s \geq \omega _{_B}$.
\par The value of $\bar B$ is somewhere in the interval from $\sim
B_{lc}$ to $\sim (c/\Omega r_i)^3 B_{lc}$, here
$$B_{lc} \simeq B^d_{_S}\left({\Omega R\over c}\right)^3
\simeq 10^4\left({B^d_{_S}\over 10^{12}\, {\rm Gauss}}
\right)\left({\Omega \over 70\,\, {\rm s}^{-1}}\right)^3
\,\, {\rm Gauss} \eqno (2)$$
is the characteristic value
of the magnetic field at the light cylinder of
pulsar, $B^d_{_S}$ is the dipole component of the magnetic field
at the neutron star surface,
$R\simeq 10^6$ cm is the neutron star radius, and $r_i$
is the distance from the center of pulsar to the inner boundary
of the outer gap.
\par The value of $r_i$ is determined by both the structure of the
magnetic field in the pulsar magnetosphere and the angle $\chi$
between the stellar magnetic moment and the spin axis.
At $R\ll r < c/\Omega$, a pure dipole field is a good
approximation for the pulsar magnetosphere. In this
approximation the value of $r_i$ is ${2\over 3}(c/\Omega)$
at $\chi = 0$. This is its maximum value;
$r_i$ drops with increasing $\chi$. For
$\chi \simeq \pi /4$ we have $r_i \simeq {1\over 2}(c/\Omega)$.
When the angle $\chi$ goes to $\pi /2$, the following
equation for $r_i$ may be used (Halpern and Ruderman  1993,
hereafter HR)
$$r_i\simeq {4\over 9}(\tan \chi)^{-2}(c/\Omega)\,. \eqno (3)$$
I will assume, in this Section, that $r_i$ is not more than a few
times smaller than $c/\Omega$, i.e. $\pi/2 - \chi$ is not too small.
This is because only for this case the outer gap model was
developed in CHRa.
\par In CHRb it was
assumed that $\bar B \simeq B_{lc}$. This is a natural
assumption because the strength of the magnetic field is $\sim B_{lc}$
in the main part of the outer gap as long as
the Vela-like pulsar is not
too near the death line. Substituting $B_{lc}$ for $\bar B$ into
equation (1), we have
$$\omega_s \simeq 10^{16}
\left({B_{_S}^d\over 10^{12}\, {\rm Gauss}}\right)^{-3}\left({\Omega
\over 70\,\,{\rm s}^{-1}}\right)^{-7}\,\,
{\rm s}^{-1}\,, \eqno (4)$$
\par In the outer gap of a Vela-like pulsar the primary
electron-positron pairs are created by the tertiary photons in a
Coulomb  field of the ultrarelativistic primary particles. The
cross section for this process is $\sigma _{_\pm}\simeq 10^{-26}$
cm$^2$ (Heitler 1954; Cheng and Ruderman 1977). In a steady state,
one electron-positron pair is created inside the outer gap
per primary particle before it leaves the outer gap, i.e.,
$$\sigma _{_\pm}n_s l \simeq 1\,, \eqno (5)$$
where $n_s$ is a gap-averaged
 tertiary photon density and $l = c/\Omega
- r_i \simeq {1\over 2}c/\Omega$ is the outer gap length.
\par The luminosity of a Vela-like pulsar in the tertiary photons
is (CHRb)
$$L_s\simeq 2 n_s\hbar \omega _s c bl\,, \eqno (6)$$
where $b$ is the outer gap breadth, $b\simeq l$.
\par From equations (1), (4), (5) and (6) we have
$$L_s \simeq { c^2\hbar \omega _s\over \Omega \sigma _{_\pm}}
\simeq 10^{34}\left({\bar B\over 10^4\, {\rm Gauss}}\right)^{-3}
\left({\Omega \over 70\, {\rm s}^{-1}}\right)\,\,\,\,
{\rm ergs\,\, s}^{-1}\simeq $$
$$\simeq 10^{34}\left({B^d_{_S}\over 10^{12}\, {\rm Gauss}}\right)^{-3}
\left({\Omega \over 70\, {\rm s}^{-1}}\right)^{-8}\,\,\,\,
{\rm ergs\,\, s}^{-1}\,.\eqno (7)$$
\par For the Vela pulsar, $B_{_S}^d\simeq 3.5\times 10^{12}$ Gauss and
$\Omega \simeq 70$ s$^{-1}$,
from equation (4) a typical frequency
of the tertiary radiation is $\omega _s \simeq 2\times 10^{14}$ s$^{-1}$,
i.e. at the IR region. The expected IR luminosity of the Vela pulsar
in tertiary photons is $\sim 10^{32}$ ergs s$^{-1}$
(see CHRb and equation (7)).
This IR luminosity would not have made Vela an IRAS source.
\bigskip
2.2 {\it Death line, death valley and pulsar luminosity in $\gamma$-rays}
\medskip
The whole outer gap can operate as a Vela-like generator of
radiation in the pulsar magnetosphere only if the pulsar period
is smaller than the following value (RC; Ho 1989; CR)
$$P_v = 0.1 \left({B^d_{_S}\over 10^{12}\, {\rm Gauss}}
\right)^{5/12}\, {\rm s}\,. \eqno (8)$$
It was predicted (RC) that the $\gamma$-ray luminosity of a
Vela-like pulsar, $L_\gamma$, increases sharply when the pulsar period,
$P$, goes to $P_v$. At $P = P_v$ the $L_\gamma$ value may be as high as
$$L^{\rm max}_\gamma \simeq \zeta \dot E_{\rm rot} \simeq
4\times 10^{35}\zeta
\left({B^d_{_S}\over 10^{12}\, {\rm Gauss}}\right)^{1/3}\,\,
\,\, {\rm ergs\,\,s}^{-1} \,, \eqno (9)$$
where (Ostriker and Gunn 1969)
$$\dot E_{\rm rot} = I\Omega \dot\Omega \simeq {2\over 3}
{(B^d_{_S})^2R^6\Omega ^4\over c^3} \eqno (10)$$
is the total spin-down power of the neutron star and $\zeta$ is
a fraction of open magnetic field lines which pass through
the outer gap. At $P\simeq P_v$ the value of $\zeta$ is $\sim 0.5$.
\par The death line of an outer magnetosphere accelerator is (CR)
$$5\log \left({B_{_S}^d\over 10^{12}\, {\rm Gauss}}\right)
- 12\log\left({P\over 1\,{\rm s}}\right) = 12 + 15\log \eta
\,, \eqno(11)$$
where $\eta = \Omega r_i/c$.
\par This line relates to the pulsar period
$$P =  P_d = 0.1 \left({B^d_{_S}\over 10^{12}\, {\rm Gauss}}
\right)^{5/12}\, \eta ^{-5/4} \,{\rm s}\,. \eqno (12)$$
Equations (11) and (12) are valid only if $\eta$ is not $\ll 1$
as it is assumed above.
\par If the pulsar period is between $P_v$ and $P_d$
(the death valley) only a part of the outer gap, from its
inner boundary, $r = r_i$, to the distance
$\sim r_i(P_d/P)^{4/5}$, can operate as a Vela-like outer gap.
In this region of the outer gap the main mechanism of energy
losses for primary particles is inverse Compton scattering on
tertiary photons.
Since the electric field component $E_\parallel$ is more or less
constant along the outer gap (CHRa)
the $\gamma$-ray luminosity from this part of the outer gap at
$P_v < P < P_d$ is
$$L_\gamma \simeq
\zeta I\Omega\dot\Omega{r_i\lbrack (P_d/P)^{4/5} -1 \rbrack
\over c/\Omega -r_i} \simeq
\zeta I\Omega\dot\Omega\left({P^{4/5}_d -
P^{4/5} \over P^{4/5}_d - P^{4/5}_v} \right) \,\,,\eqno (13)$$
where the dimensionless parameter $\zeta$ is
$$\zeta \simeq 1 - (P/P_d)^{4/5}\,\,. \eqno (14)$$
\par The total $\gamma$-ray luminosity of a Vela-like pulsar
inside the death valley may be more than the value given
by equations (13) and (14). The point is that
in the region from $\sim r_i(P_d/P)^{4/5}$ to the light cylinder
the primary particles are accelerated by $E_\parallel$ and radiate
the curvature $\gamma$-rays.
\par If the pulsar period is
$$P < P_c \simeq
2\pi\left( {e^5(B_{_S}^d)^3R^9\over 24 m^4c^{15}}
\right)^{1/7}\simeq 0.6\left({B_{_S}^d\over 10^{12}\,\,
{\rm Gauss}}\right)^{3/7}\,\,\,{\rm s}\,, \eqno (15)$$
the primary particles radiate a main part of their energy by the
curvature $\gamma$-rays before reaching the light cylinder. In this
case the pulsar luminosity in $\gamma$-rays inside the death valley
may be as high as
$$L^{\rm max}_\gamma \simeq \zeta I\Omega\dot\Omega\,, \eqno (16)$$
where $\zeta$ is determined by equation (14).

\bigskip
2.3 {\it Tertiary radiation near the death line}
\medskip
{}From equation (4) we can see that $\omega _s$ increases
sharply with decreasing $\Omega$.
At the boundary of the death valley, $P\simeq P_v$, from equations
(4) and (8) the typical frequency of the tertiary photons is
$$\nu _s ={\omega _s \over 2\pi}
\simeq 3\times 10^{15}\left({B^d_{_S} \over
10^{12}\,{\rm Gauss}}\right)^{-1/12}\,\,\,\,{\rm Hz}\,\,.\eqno (17)$$
Hence, in the outer gap model of CHRb it is expected that
when a Vela-like pulsar in the process of deceleration
of its rotation goes to the death valley the typical frequency of the
tertiary photons goes from IR range for the Vela pulsar to
the optical range and to the UV range more.
\par From equations (7) - (9) the expected luminosity of a Vela-like
pulsar in the optical and UV ranges at $P\simeq P_v$ is
$$L_{\rm opt\,+\,UV}
\simeq 2\times 10^{34} \left({B^d_{_S}\over 10^{12}\, {\rm Gauss}}
\right)^{1/3}\,\,\,\,{\rm ergs\,\,s}^{-1}\simeq 0.1 L^{\rm max}_\gamma
\,,\eqno (18)$$
i.e., near the death valley not only the $\gamma$-ray emission
of a Vela-like pulsar has to increase as it was predicted
by CR but its optical and UV emission has to increase as well.
\par The tertiary photon spectrum has the following
frequency dependence (Bekefi 1966; CHRb)
$$I_\nu \propto \cases{(\nu /\nu_s)^{1/3}&at $\nu \ll\nu _s\,$,\cr
(\nu/\nu _s)^{1/2}{\rm exp}\,
\lbrack - (\nu /\nu _s)\rbrack&at $\nu\gg\nu _s
\,.$\cr}
\eqno (19)$$
Therefore, the luminosity of a Vela-like pulsar at $P\simeq P_v$ in
the optical range, between $4\times 10^{14}$ Hz and $7.5 \times
10^{14}$ Hz, is
$$L_{\rm opt} \simeq \left({7.5\times 10^{14} \,\,{\rm Hz}\over
\nu _s}\right)^{4/3}L_{\rm opt + UV} \simeq3\times
10^{33}\left({B^d_{_S}\over 10^{12}\,\,{\rm Gauss}}\right)^{4/9}
\,\,\, {\rm ergs\,\,s}^{-1}\,.\eqno (20)$$
\par For a Vela-like pulsar inside the death valley
the magnetic field at the distance
$\sim r_i(P_d/P)^{4/5}\simeq (c/\Omega)(P_v/P)^{4/5}$ from the pulsar
may be taken as a gap-averaged magnetic field
$$\bar B \simeq B_{lc}\left({P\over P_v}\right)^{12/5}\,. \eqno (21)$$
In this case from equations (1), (2), (8) and (21) we have
$$\nu _s\simeq 3\times 10^{15}\left({\Omega \over 70\,\,{\rm s}^{-1}}
\right)^{1/5}\,\,\,\,{\rm Hz}\,.\eqno (22)$$
{}From this equation we can see that the typical frequency of the
tertiary photons does not change inside the death valley
essentially and it lies in the optical and UV ranges.
\par The pulsar luminosity in the tertiary
photons inside the death valley may be estimated from equation (6)
in which the outer gap length $l$ is $\sim r_i\lbrack (P_d/P)^{4/5}
- 1\rbrack$. Using this, from equations (8), (14) and (16) we can get
$${L_{\rm opt + UV}\over L^{\rm max}_\gamma} \simeq
{3h\nu _sc^5\over \sigma _{_\pm}(B^d_{_S})^2R^6\Omega^5}
\left({P_v\over P}\right)^{4/5} \simeq 0.1 \left({P\over P_v}
\right)^4\,\,.\eqno (23)$$
A pulsar with so high ratio of $L_{\rm opt\,+\,UV}$ and $L_\gamma$
has to be observed in the optical and UV ranges if it is
observed as a $\gamma$-ray source.
\par There is a very essential assumption which is used above,
namely: like CHRb we assumed that a gap-averaged magnetic field
$\bar B$ is approximately equal to the magnetic field at the
outer boundary of the outer gap.
If $\bar B$ is higher than the value we have used, both the
typical frequency of the tertiary photons and the pulsar
luminosity in these photons decrease.
\par In fact, in the outer gap model of CHRb
there is no prescription for estimating $\bar B$
more or less exactly. It is known only that $\bar B$ is somewhere
between $B_{lc}$ and $(c/\Omega r_i)^3B_{lc}$. If the angle $\chi$
is small, we have $r_i \simeq {2\over 3}(c/\Omega)$ and
$\bar B < (3/2)^3B_{lc}$. Using this from equations
(1), (6) - (9), (17) and (18) we can see that $\nu _s >
3(2/3)^9\times 10^{15}{\rm Hz}\simeq 10^{14}$ Hz and
$L_s/L_\gamma > 3\times 10^{-3}$. In this case
the ratio of the optical and $\gamma$-ray luminosities is
$${L_{\rm opt}\over L_\gamma} \simeq {L_s\over L_\gamma}
\left({4\times 10^{14}\,\,{\rm Hz} \over \nu _s
}\right)^{3/2}{\rm exp}\left[ -{4\times 10^{14}\,\,{\rm Hz}
\over \nu _s}\right] > 4\times 10^{-4}\,. \eqno (24)$$
Optical emission from a $\gamma$-ray source at energies
$\sim 10 -10^4$ MeV has to be detected if its flux of energy
in the optical range is
of the order of or more than $\sim 10^{-6}$ times the flux of energy
in $\gamma$-rays. Hence,
optical emission must be observed from a Vela-like pulsar
inside (or near) the death valley if the angle $\chi$ is small
enough.
\par If $\chi > \pi /4$, i.e. $r_i < {1\over 2}
(c/\Omega)$, the low limit on $\nu _s$ at $P\simeq P_v$
is equal or smaller than $\sim 3\times 2^{-9}\times 10^{15}$
Hz $\sim
10^{13}$ Hz and the flux of optical emission
from the pulsar may be very small. But, if a Vela-like pulsar
goes to the death line the magnetic field values at both the outer
boundary of the outer gap and its inner boundary draw
near each other. In this case equation (22) gives a more
and more exact estimate of $\nu _s$, $\nu _s \sim 10^{15}$ Hz.
Therefore, the optical emission has to be
detected from a Vela-like pulsar if the pulsar is close enough to the
death line and its inclination angle $\chi$ is not too great.
This result is independent of how $\bar B$ is determined.

\heading{3. NEUTRON STAR HEATING BY REVERSED PARTICLES AND
X-RAY EMISSION}
A part of energy which is released in the outer gaps of
pulsars is transferred to the neutron star surface by the
ultrarelativistic primary particles.
Let us estimate the energy flux which is deposited onto the neutron star
surface by both the reversed ultrarelativistic primary particles
and their radiation.
\par The potential drop
$\Delta V$ across an outer gap may be as high as (CHR, HR)
$$\Delta V_{\rm max} = {\Omega ^2B^d_{_S}R^3\over 2c^2} \simeq
6.6\times 10^{14}\left({B^d_{_S}\over 10^{12}\,\,{\rm Gauss}}\right)
\left({P\over 0.1\,\,{\rm s}}\right)^{-2}\,\,\,\, {\rm V}
\,. \eqno (25)$$
In the case of the Vela pulsar the energy of the reversed primary
particles which move from the outer gaps to the pulsar surface
is a few orders smaller than $e\Delta V_{\rm max}$ (CHRb). This is
because of very
high energy losses of primary particles in the process of their
inverse Compton scattering on the tertiary photons. But when a
Vela-like pulsar goes to the death valley the energy losses of primary
particles because of inverse Compton scattering drop (CHRb; RC).
For a Vela-like
pulsar near the death line the energy of primary particles
at the inner boundary of the outer gap is smaller but not
more than a few times than $e\Delta V$.
\par Curvature radiation is the main mechanism of energy losses
for primary particles during their motion from the inner boundary
of the outer gap to the neutron star surface.
Let us assume that the magnetic field of the neutron star is dipole.
As it is noted above this is a good approximation for the magnetic field
structure at a distance more
than a few $R$ but smaller than the radius of the light cylinder.
In this case if a primary particle is starting
from the distance $r_i$ to the pulsar surface
along the last closed $\bf B$ field line then its energy at the
distance $r$ is (Ochelkov and Usov 1980; Harding 1981)
$$\gamma (r) = \gamma _{_0}
\left[ 1+ {9\over 8}{e^2\Omega \gamma _{_0}^3\over
mc^3}\ln\left({r_i\over r}\right)
\right] ^{-1/3}\,, \eqno (26)$$
where $\gamma _{_0}$ is the Lorentz factor of primary particle at the
distance $r_i$.
\par In the case $\gamma ^3 \ll \gamma _{_0}^3$ which is relevant to
Vela-like pulsars from equation (26) we have
$$\gamma (r)\simeq \left[ {9\over 8}{e^2\Omega\over
mc^3}\ln\left({r_i\over r}\right)\right] ^{-1/3}\,, \eqno (27)$$
\par Primary particles begin to irradiate the neutron star
surface when $r < r_*$ (see Fig. 1), where
$$r_* \simeq R\left({4c\over \Omega R}\right)^{1/3}\,. \eqno (28)$$
Curvature photons which are generated by primary
particles during their motion to the neutron star surface are
in the $\gamma$-ray range, $\hbar \omega \simeq (3/2)(\hbar c/R_c)
\gamma ^3\sim 10^3- 10^5$ MeV
$\lbrack$ here $R_c \simeq (rc/\Omega)^{1/2}$ is the
curvature radius of the magnetic field lines $\rbrack$. These photons
are absorbed in the magnetic field near the pulsar with electron-positron
pair creation. All energy of secondary pairs and main part of
energy of their synchrotron radiation is deposited onto the
neutron star surface. This energy is reradiated  by the neutron
star surface into X-rays.
\par The energy which is reradiated in X-rays per one primary
particle is $\sim \gamma (r_*) mc^2$. Since the energy which
is radiated in $\gamma$-rays by a primary particle can not be
more than $e\Delta V_{\rm max}$ we have the following lower limit
on the ratio of the X-ray and $\gamma$-ray luminosities of a Vela-like
pulsar near (or inside) the death valley:
$${L_x\over L_\gamma} > {\gamma (r_*)mc^2\over e\Delta
V_{\rm max}}\simeq \left[ {9\over 8} {e^2\Omega\over mc^3}
\ln \left({r_i\over r_*}\right)\right] ^{-1/3}
{mc^2\over e\Delta V_{\rm max}}\simeq $$
$$\simeq 0.9\times 10^{-2}\left({B^d_{_S}\over 10^{12}\,\,{\rm Gauss}}
\right)^{-1}\left({P\over 0.1\,\,{\rm s}}\right)^{7/3}\left[
\ln \left({r_i\over r_*}\right)\right] ^{-1/3}\,.\eqno (29)$$
A Vela-like pulsar with $B^d_{_S}\simeq 10^{12}$ Gauss and
$P > 0.1$ s has $L_x/L_\gamma > 10^{-2}$. Pulsars
with so high ratio $L_x/L_\gamma$
have to be observed as a X-ray source if they are observed
as a $\gamma$-ray source.

\heading{4. GEMINGA AS A CANDIDATE FOR VELA-LIKE PULSARS}
\par For Geminga the neutron star parameters are $P = 0.237$ s
and $B^d_{_S}\simeq 1.6\times 10^{12}$ Gauss (Halpern and Holt
1992; Bertsch et al. 1992). The distance to Geminga is unknown;
the upper limit on the distance is $\sim 400$ pc.
The $\gamma$-ray luminosity of Geminga
may be as high as the total spin-down power, $L_\gamma \sim
I\Omega\dot\Omega$, and its period is between $P_v$
and $P_d$. Therefore, Geminga could be a Vela-like pulsar
inside the death valley (HR). In this case,
from the observed phase separation
between two $\gamma$-ray pulses  it follows that $r_i <
0.05(c/\Omega)\simeq 5\times 10^7$ cm and $\chi > 70^o$ (HR).
\par From equations (28) and (29) the ratio $L_x/L_\gamma$
which is expected for Geminga in the outer gap model (CHRb) is
more than
$4\times 10^{-2}$. This is in contradiction with the available data,
according to which Geminga's X-ray luminosity is $\sim (1 - 2)
\times 10^{-3}
L_\gamma$. Even if we take into account only the heating of the polar
caps by the reversed primary particles, from equation (29) with
$r_i = R$ we have $L_x/L_\gamma > 2\times 10^{-2}$ which is
an order more than the observed ratio $L_x/L_\gamma$ for Geminga.
\par Since $\eta _i < 0.05 \ll 1$ the results of Section 2
can not be applied to Geminga to get any reliable estimate of
its optical luminosity. However,
these results may be used to get some restrictions on the outer
gap parameters.
\par Taking into account that the optical luminosity of Geminga
in the tertiary radiation
is not more than $\sim 10^{-6}L_\gamma$ from equations (6), (7),
(14), (16) and (19) we have
$${L_s\over L_\gamma}\left({4\times 10^{14}\,\,{\rm Hz}\over
\nu _s}\right)^{3/2}{\rm exp} \left(-{4\times 10^{14}\,\,{\rm Hz}
\over  \nu _s}\right) < 10^{-6}\eqno (30)$$
or
$$\nu _s < 2.7\times 10^{13}\,\,{\rm Hz}\,. \eqno (31)$$
{}From equations (1) and (31) it is followed that
$$B > 2\times 10^4\,\,{\rm Gauss}\,.\eqno (32)$$
If the outer gap region with $B < 2\times 10^4$ Gauss is irradiated
by the tertiary photons so that it operates effectively as a
Vela-like generator of $\gamma$-rays then the optical luminosity
from the outer gap has to be more than $\sim 10^{-6} L_\gamma$.
\par On the other hand, the outer gap region with $B > 6\times
10^4$ Gauss can not operate as a Vela-like generator of both
$\gamma$-rays and primary particles. Indeed, in such a strong
magnetic field the main mechanism of the energy losses for
primary particles is the curvature radiation but not Compton scattering
(CHRb). Therefore, at $B > 6\times 10^4$ Gauss the sequence of
processes which was assumed by CHRb for a Vela-like outer gap
(see Section 2) breaks down.
\par The magnetic field varies from $\sim B_{lc}\simeq 0.9\times
10^3$ Gauss to $\sim (c/\Omega r_i)^3B_{lc} >$ $7\times
10^6$ Gauss along the outer gap of Geminga. Only a small region
of the outer gap with $B$ in the range from $\sim 2\times 10^4$
Gauss to $\sim 6\times 10^4$ Gauss can operate as a Vela-like
generator of $\gamma$-rays. The length of this region along
${\bf B}$ is
$$\Delta r = r_1 - r_2\simeq 0.1(c/\Omega)\simeq 0.1 l\,,
\eqno (33)$$
where
$$r_1\simeq {c\over \Omega} \left({B_{lc}\over 2\times 10^4
\,\,{\rm Gauss}}\right)^{1/3} \,\,\,\,\,\, {\rm and}\,\,\,\,\,\,\,
r_2\simeq {c\over \Omega}\left({B_{lc}\over 6\times 10^4
\,\,{\rm Gauss}}\right)^{1/3} \eqno (34)$$
are the distance from the center of the neutron star to the upper and
inner boundaries of the Vela-like region, respectively.
\par Since $\Delta r \ll l$ the potential drop
$\Delta V_{\Delta r}$
on the Vela-like region of the outer gap is small,
$\Delta V_{\Delta r}\ll \Delta V_{\rm max}$. The value of
$\Delta V_{\Delta r}$ may be estimated roughly as
$$\Delta V_{\Delta r}\simeq {\Delta r\over l}\Delta V_{\rm max}
\simeq 0.1 \Delta V_{\rm max}\,. \eqno (35)$$
For Geminga, $\Delta V_{\rm max}\simeq 2\times 10^{14}$ V,
the primary particles may be accelerated in the Vela-like
region of the outer gap up to the energy $\sim 2\times 10^{13}$ eV.
This energy is enough for pair production in the process of primary
particle interaction  with the tertiary photons of which the energy
is $< 0.1$ eV. Therefore, this region, in principle, can
operate as a Vela-like generator of $\gamma$-rays and particles.
But, the location of this region between $r_2\simeq 0.25
(c/\Omega)$ and $r_1\simeq 0.35 (c/\Omega)$ is in contradiction
with the result of HR that for Geminga the region of $\gamma$-ray
generation is at the distance $r_e < 0.05 (c/\Omega)$.
\par The period of Geminga is smaller than $P_c$ (see equation (15)).
Therefore, the motion of primary particles between the upper boundary
of the Vela-like region, $r = r_1$, and the light cylinder, $r =
c/\Omega$, is quasi-stationary, i.e. the energy losses of particles
$(2/3)(e^2c\gamma ^4/R_c^2)$ because of the curvature radiation are
equal to the energy $eE_\parallel c$ which particles get from the
electric field per unit time.
In this case the Lorentz factor of particles is
$$\gamma \simeq \left({3\over 2}{E_\parallel R^2_c\over e}\right)^{1/4}
\,.\eqno (36)$$
The mean energy of the curvature photons which are generated by these
particles is
$$\bar \epsilon _\gamma = {3\over2}{\hbar c\over R_c}\gamma ^3 \simeq
{3\over 2}\hbar c \left({3\over 2}{E_\parallel \over e}\right)^{3/4}
R_c^{1/2}\,.\eqno (37)$$
For Geminga, $E_\parallel \simeq \Delta V_{\rm max}/l \simeq 6\times
10^2$ e.s.u. and $R_c\simeq 10^9$ cm, from equation (37) we have
$\bar \epsilon _\gamma \simeq 10^3$ MeV. This value of $\bar \epsilon_
\gamma$ is suitable to explain the $\gamma$-rays from Geminga.
\par The $\gamma$-ray luminosity of Geminga at $\sim 10 - 10^4$ MeV is
$$L_\gamma (10 - 10^4\,\,{\rm MeV})\simeq {(c/\Omega) - r_2 \over
\Delta l}L_1 \simeq 7.5 L_1\,,\eqno (38)$$
where $L_1$ is the power which is released by the primary
particles in the Vela-like region of the outer gap.
\par Since the $\gamma$-ray luminosity of the Vela-like region can not be
more than $L_1$, the main mechanism of $\gamma$-ray generation for
Geminga is curvature radiation. This differs essentially from
Vela-like pulsars for which the synchrotron radiation was proposed in
CHRb as a mechanism of $\gamma$-ray generation at $\sim 10 - 10^4$
MeV. Hence, Geminga is not a close relative of Vela-like pulsars.
\par A very strong radiation in TeV $\gamma$-rays
is expected from Geminga in the outer gap model of CHRb.
The point is that in the model of
a Vela-like pulsar the energy of primary particles before being
transferred into $\gamma$-rays  with energies $\epsilon _\gamma
\sim 10 - 10^4$ MeV
is in very high energy $\gamma$-rays, $\epsilon _\gamma \sim
10^{12} - 10^{13}$ eV (see CHRb and Section 2). Near the death line
the mean free path of very high-energy $\gamma$-rays for pair production
in collisions with the tertiary photons is more or less equal to
the outer gap dimensions. Therefore, the absorption of very high
energy $\gamma$-rays is not too strong and an essential part of these
$\gamma$-rays has to escape  from the outer gap. The $\gamma$-ray
luminosity at TeV energies for a Vela-like
pulsar near the death line is (Cheng and de Jager 1990)
$$L_\gamma ({\rm TeV})\simeq {1\over 2}L_1\,,\eqno (39)$$
In the case of modified outer gap model in which the Vela-like
region is a source of primary particles and the main part of
$\gamma$-rays is the curvature radiation,
from equations (38) and (39) we have
$${L_\gamma ({\rm TeV})\over L_\gamma (10 - 10^4\,\,{\rm MeV})}
\simeq {\Delta l \over 2\lbrack (c/\Omega ) - r_2 \rbrack}\,.
\eqno (40)$$
For Geminga the expected ratio $L_\gamma ({\rm TeV})/L_\gamma
(10 - 10^4\,\,{\rm MeV})$ is $\sim 0.07$, which might be uncertain
within a factor of 2 or so. The upper limit on the $\gamma$-ray flux
from Geminga at TeV energies, $\epsilon _\gamma \sim 10^{12} - 10^{13}$
eV, is $\sim 10^{-10}$ ergs s$^{-1}$ cm$^{-2}$ (Weekes 1988; Akerlof
et al. 1993 and references therein). Since the
observed $\gamma$-ray flux at $\sim 10 - 10^4$ MeV is at least
$\sim 2.4\times
10^{-9}$ ergs s$^{-1}$ cm$^{-2}$ (Fichtel et al. 1975; Halpern and
Holt 1992), the upper limit on the ratio $L_\gamma ({\rm TeV})/
L_\gamma (10 - 10^4\,\,{\rm MeV})$ is $\sim 0.04$. We can see that
a contradiction between the $\gamma$-ray luminosity at TeV energies
and the available data on the TeV $\gamma$-rays begins to show for
Geminga even in the modified outer gap of Geminga.
\par As to the case when the whole of the outer gap operates as a
Vela-like generator of $\gamma$-rays, $L_1\simeq I\Omega\dot
\Omega$, it is expected that $L_\gamma ({\rm TeV})\simeq
L_\gamma (10 - 10^4\,\,{\rm MeV})$ (RC; Cheng and de Jager 1990).
This kind of an outer gap model is ruled out for Geminga.

\heading{5. CONCLUSIONS AND DISCUSSION}
I have considered in this paper the radiation of Vela-like
pulsars near the death line. It is shown that X-ray emission
with $L_x/L_\gamma > 10^{-2}$ is expected from such a
pulsar. Besides, the optical emission of a Vela-like pulsar
has to increase sharply if the pulsar is close enough to the death
line and its inclination angle is not too great.
The optical luminosity $L_{\rm opt}$ may be up to
$\sim 10^{-2}L_\gamma$. If $L_{\rm opt} \ll 10^{-2} L_\gamma$,
a distinguishing feature of the tertiary optical emission is a
very steep spectrum.
\par Using these results, Geminga is considered as a candidate for a
Vela-like pulsar. It is shown that the outer gap model of CHRb is
at odds with the available data on Geminga.
The outer-gap model of Geminga in which
the main part of the outer gap volume operates as a Vela-like
generator of $\gamma$-rays is ruled out both by the upper limit
on the flux of the tertiary photons in optical range, and also by
the upper limit on the $\gamma$-ray flux from Geminga at TeV
energies. A Vela-like mechanism of $\gamma$-ray generation
can operate only in a compact region for which the
length along the magnetic field is essentially smaller than the
outer gap length. In principle, this compact Vela-like region
of the outer gap can be a source of primary particles.
The main mechanism of $\gamma$-ray
generation in the Geminga magnetosphere is the curvature radiation
of primary particles
but not the synchrotron radiation as it was assumed in CHRb for
a Vela-like outer gap.
Summarizing, we can conclude that Geminga is not a
close relative of Vela-like pulsars. This does not mean that a general
idea of HR on an outer gap model is ruled out for Geminga. It means only
that if an outer gap model is developed for Geminga, it
has to differ qualitatively
from the outer gap model of CHRb for Vela-like pulsars.
\par Harding, Ozernoy and Usov (1993) have considered the polar-gap
model for Geminga which is an alternative to an outer-gap model. They
have suggested that Geminga is more or less a typical pulsar, as
has been proposed by Ozernoy and Usov (1977) for unidentified
high-energy $\gamma$-ray sources.
In this model the $\gamma$-ray emission of Geminga is the curvature
radiation of the primary particles which are accelerated near the
pulsar surface in the polar gaps. The polar gap model can
describe the X-ray and $\gamma$-ray radiation of Geminga
if the distance to Geminga is not more than $\sim 30 - 40$ pc.
\par A strong IR emission was predicted from the Vela pulsar (see
CHRb and above). The observation of this emission
would be a direct confirmation of the outer gap model of CHRb. The other
way to verify the outer gap model of CHRb for a Vela-like pulsar is the
observation of the optical tail of the tertiary radiation with a
very steep spectrum. In
connection with this, it is worth noting that the $\gamma$-ray pulsar
PSR 1706-44 is very promising for such a kind of observations.
Indeed, for PSR 1706-44 the neutron star parameters are
$P = 0.102$ s and $B_s\simeq 3\times 10^{12}$ Gauss (Thompson et al.
1992), and the
typical frequency of the tertiary photons for this pulsar is $\sim 4$
times more than for the Vela pulsar (see equation (4)). This is very
essential for observation of the optical tail because of the
exponential dependence of the tertiary photon spectrum
at high frequencies, $\nu \gg \nu_s$ (see equation (19)).

\heading{FIGURE CAPTION}
{\it Figure 1.} Irradiation of the neutron star surface
by the curvature $\gamma$-ray emission of the reversed primary
particles. At $r\leq r_*$ practically all radiation of the particles
is deposited onto the neutron star surface. ${\bf \mu}$ is the
stellar magnetic moment. $B$ denotes the last closed ${\bf B}$
field line. $R$ is the radius of the neutron star.

\heading{ACKNOWLEDGMENTS}
It is a pleasure for me to thank M. Milgrom for many valuable
discussions.

\heading{REFERENCES}
Akerlof, C.W., et al. 1993, Astron. Ap., {\bf 274}, L17

Bekefi, G. 1966, Radiation Processes in Plasmas (NY: J. Wiley and Sons)

Bertsch, D.L., et al. 1992, Nature, {\bf 357}, 306

Chen, K., and Ruderman, M.A. 1993, ApJ, {\bf 402}, 264 (CR)

Cheng, A., and Ruderman, M.A. 1977, ApJ, {\bf 214}, 598

Cheng, A., Ruderman, M.A., and Sutherland, P.G. 1976, ApJ, {\bf 203},
209

Cheng, K.S., and de Jager, O.C. 1990, Nuclear Phys., {\bf 14A}, 28

Cheng, K.S., Ho, C., and Ruderman, M.A. 1986a, ApJ, {\bf 300}, 500 (CHRa)

Cheng, K.S., Ho, C., and Ruderman, M.A. 1986b, ApJ, {\bf 300}, 522 (CHRb)

Fichtel, C.E. et al. 1975, ApJ, {\bf 196}, 163

Halpern, J.P., and Holt, S.S. 1992, Nature, {\bf 357}, 222

Halpern, J.P., and Ruderman, M.A. 1993, ApJ (in press) (HR)

Harding, A.K. 1981, ApJ, {\bf 245}, 267

Harding, A.K., Ozernoy, L.M., and Usov, V.V. 1993, MNRAS (in press)

Heitler, W. 1954, Quantum Theory of Radiation (Oxford Univ. Press)

Ho, C. 1989, ApJ, {\bf 342}, 396

Holloway, N.J. 1973, Nature  Phys. Sci., {\bf 246}, 6

Krause-Polstorff, J., and Michel, F.C. 1985a, Astron. Ap., {\bf 144}, 72

Krause-Polstorff, J., and Michel, F.C. 1985b, MNRAS, {\bf 213}, 43

Ochelkov, Yu.P., and Usov, V.V. 1980, Ap. Sp. Sci., {\bf 69}, 439

Ostriker, J.P., and Gunn, J.E. 1969, ApJ, {\bf 157}, 1395

Ozernoy, L.M., and Usov, V.V. 1977, SvA, {\bf 21}, 425

Ruderman, M.A., and Cheng, K.S. 1988, ApJ, {\bf 335}, 306 (RC)

Thompson, D.J. et al. 1992, Nature, {\bf 359}, 615

Weekes, T.C. 1988, Physics Reports, {\bf 160}, 1

\bye